\documentclass{article}

\usepackage{PRIMEarxiv}
\usepackage{tabularx}
\usepackage{amsmath}
\usepackage[utf8]{inputenc} 
\usepackage[T1]{fontenc}    
\usepackage{hyperref}       
\usepackage{url}            
\usepackage{booktabs}       
\usepackage{amsfonts}       
\usepackage{nicefrac}       
\usepackage{microtype}      
\usepackage{lipsum}
\usepackage{fancyhdr}       
\usepackage{graphicx}       
\graphicspath{{media/}}     

\title{Sound Terminology Describing Production and Perception of Sonification
\thanks{
\textbf{ }} 
}

\author{
  Ziemer, Tim\\
  Institute of Systematic Musicology \\
  University of Hamburg \\
  Germany\\
  \texttt{tim.ziemer@uni-hamburg.de}
}

\begin{document}
\maketitle

\begin{abstract}
Sonification research is intrinsically interdisciplinary. Consequently, a proper documentation of, and interdisciplinary discourse about a sonification is often hindered by terminology discrepancies between involved disciplines, i.e., the lack of a common sound terminology in sonification research. Without a common ground, a researcher from one discipline may have troubles understanding the implementation and imagining the resulting sound perception of a sonification, if the sonification is described by a researcher from another discipline. To find a common ground, I consulted literature on interdisciplinary research and discourse, identified problems that occur in sonification, and applied the recommended solutions. As a result, I recommend considering three aspects of sonification individually, namely 1.) Sound Design Concept, 2.) Objective and 3.) Method, clarifying which discipline is involved in which aspect, and sticking to this discipline's terminology. As two requirements of sonifications are that they are a) reproducible and b) interpretable, I recommend documenting and discussing every sonification design once using audio engineering terminology, and once using psychoacoustic terminology. The appendix provides comprehensive lists of sound terms from both disciplines, together with relevant literature and a clarification of often misunderstood and misused terms.
\end{abstract}


\section{INTRODUCTION}
While early definitions refer to sonification as ``the use of non-speech audio to convey information'' \cite{report}, newer definitions add the requirements of the sound being a) \emph{reproducible} and b) \emph{interpretable} \cite[p. 274]{arbitrary}\cite{taxonomy}. This paper is supposed to help sonification researchers meet these requirements by suggesting a common sound terminology for interdisciplinary sonification research and discourse. 

Sonification is intrinsically interdisciplinary. For example, \cite{shintro} present the interdisciplinary cycle of sonification, including audio engineering, psychoacoustics and music, but also cognitive sciences, acoustics/computer music, computer science, linguistics/philosophy, social sciences, product design, data mining/statistics and many more. \cite{shintro} state that physics, acoustics, psychoacoustics, perceptual research, sound engineering and computer science are ``required to comprehend and carry out successful sonification'', while respecting that sonification also has facets concerning psychology, musicology, cognitive sciences, linguistics, pedagogies, social sciences and philosophy. The Sonification Handbook \cite{sonihb} contains chapters dedicated to psychoacoustics \cite{carlile}, perception and cognition \cite{neuortho}, human-computer interaction \cite{arbitrary}, design and aesthetics \cite{aesthetics}. David Worrall's Sonification Design book \cite{davidbuch} covers philosophy, computer science, cognitive sciences, neuroscience, psychology, psychoacoustics, and music theory. Many papers deal with the interdisciplinary nature of sonification research \cite{shintro,intson1,intsonrev,doom}.

Interdisciplinarity is often considered a strength of sonification research due to the potential to understand the overall picture rather than illuminating single fragments \cite{shintro,usepsycho}. But hurdles have been recognized for interdisciplinarity to unfold in sonification research, due to 
\begin{enumerate}
    \item no common terminology, especially concerning sound \cite{shintro,intson1,doom}
    \item blurred separation of disciplines' methods and goals \cite{shintro,davidbuch,intsonrev,doom}. 
\end{enumerate}
%
\cite{crafting} stated that promoting interdisciplinarity in sonification research has been addressed rather hesitantly. Fortunately, studies on interdisciplinary research have already identified related, general issues, and proposed respective solutions. In this paper, I relate the sonification-specific problems to general problems in interdisciplinary research and then transfer the proposed, general solutions to sonification research. The result is a list of recommendations concerning the use of sound terminology and the clarification of involved disciplines' methods and goals. The remainder of this paper is structured as follows: Sect. \ref{method} starts with a review of general issues in interdisciplinary research and proposed solutions. In Sect. \ref{intersonic} I relate these issues to sonification research, based on prominent examples. Sect. \ref{solution} aims at applying the proposed solutions to sonification research, formulating explicit recommendations to sonification researchers. These are discussed in Sect. \ref{discussion}, followed by a conclusion in Sect. \ref{conclusion}. The appendix provides literature recommendations for lexical study of sound terminology, a comprehensive list of terms and definitions describing the sound of sonifications with the vocabulary of audio engineering and of psychoacoustics, and reveals important distinctions between terms that are often confused or misused in the sonifiction literature.



%
%


\section{METHOD}
\label{method}
To tackle the issue of 1.) no common sound terminology and 2.) blurred separation of disciplines' methods and goals in sonification, I consult literature on interdisciplinary research in general to 
\begin{itemize}
    \item identify general problems in interdisciplinary research, 
    \item identify proposed solutions for interdisciplinary research in general,
    \item  relate these problems to sonification research, and, finally,
    \item apply the proposed solutions to sonification research.
\end{itemize}



\subsection{Interdisciplinarity in Research}
\label{interdisciplinarity}
\begin{table*}[!tb]
\tabcolsep8.1pt
\caption{Problems and solutions in interdisciplinary research.}
\label{tab:prob}
\begin{tabular}{lll}
\# & Problem & Solution\\
\hline
1  & Conflicting Paradigms & Name concepts, practices, and objectives\\
2  & Strong Interdisciplinarity        & Name identity criteria\\
3  & Communication Discrepancies       & \parbox{11cm}{Carry out lexical study} \\
4  & Interdisciplinary Semantic Drifts & Name discipline and stick to its terminology        \end{tabular}
\end{table*}
Literature on interdisciplinarity is often dedicated to typologies and taxonomies \cite{intertax}. An overview of interdisciplinary research is given in \cite{intertrans}.

\cite{interdis} describe the issues of conflicting paradigms. Conflicting paradigms refer to different viewpoints and interests. For example, one discipline may be interested in the object under investigation, while the other may be interested in its impact on the individual or the society. As an approach to a solution, writers and readers need to detect and interpret involved disciplines and their concepts, practices and objectives consciously.

\cite{intertrans} distinguish between ``weak'' and ``strong  interdisciplinarity''. In the case of weak interdisciplinarity, the disciplines share a ``proximity to the interests of recognition, which are reflected in pre-discursive consent, terminology, etc.'' As an example, they describe how electrical engineering and mechanical engineering may be necessary to construct a robot, where we have electric circuits on the one hand and mechanics on the other. In the case of strong interdisciplinarity, it is doubtful whether disciplines agree on the object of interest, methods and measures, and terminology. As a solution, the authors state that ``disciplines have to organise themselves within interdisciplinary research frameworks'' \cite[p. ix]{intertrans}. This means, clear identity criteria have to be named, e.g., which discipline is responsible for what, and whose methods and terminology is applied when\cite[sect. 2.3.3.2]{intertrans}.

\cite{interdis} discuss the issue of communication discrepancies. Communication discrepancies refer to cases in which, e.g., one discipline uses terms colloquially, which are specifically defined in another discipline. This can lead to the loss of detail or connotation, or even to a total misconception. As a solution, they recommend lexical studies to become aware of discipline-specific meanings and connotations of terms.

\cite{drift} describe how concepts from one discipline are borrowed by another discipline and then acquire new meanings or connotations, called \emph{Interdisciplinary Semantic Drifts} (ISDs). In contrast to the communication discrepancies discussed above, ISDs includes terms which are not used colloquially in either discipline. Still, as long as the reader does not know which discipline's definition of a term is applicable, misunderstandings can appear. The easiest solution to this issue is to name the respective discipline whose terminology you use, to stick to it as exactly as possible, and to mention whenever you include the terminology from another discipline.

Note that these issues are not completely independent of one another, but partly point in the same direction. Still, it is helpful to transfer them to interdisciplinary sonification research one by one. In summary, we can identify $4$ problems and approaches towards solutions in interdisciplinary research as listed in Table \ref{tab:prob}.

How these problems relate to sonification research is discussed in the following section, subdivided into the four raised problems.


\subsection{Interdisciplinarity in Sonification Research}
\label{intersonic}
The preceding section dealt with interdisciplinary research in general. This section shows how these problems relate to sonification research.

\subsubsection{Conflicting Paradigms}
Many sonification studies reveal relevant discipline(s) already in the title, like art \cite{arts1,arts2}, music \cite{nomusic}, speech \cite{vocaleeg}, design \cite{aesthetics,designp}, aesthetics \cite{aesthetics}, perception \cite{neuortho,davidbuch}, psychoacoustics \cite{carlile,pampas,psysoni,cars,acta,icad2019}, cognition \cite{neuortho}, mathematics \cite{vocaleeg}, philosophy \cite{artdesignphilosophy}, and/or signal processing \cite{icad2019}. This informs readers about discipline(s) involved. Unfortunately, many of these studies do not point out whether the discipline's concepts, objects of interest, or methods and measures are being applied. This gives rise to conflicting paradigms, as it makes a big difference, e.g., whether a musical sonification applies music theory and composition techniques to communicate data to a user, or whether data is used as a foundation for a music composition with the objective of an enjoyable listening experience. Furthermore, some studies' titles \cite{musicparticle,earthsing} 
include musical terms, even though neither the sonification design concept nor the  object of interest is musical. These examples of conflicting paradigms are closely related to the general issue of a blurred separation of disciplines' methods and goals \cite{shintro,davidbuch,intsonrev,doom}.






\subsubsection{Strong Interdisciplinarity}
As stated above, sonification is intrinsically interdisciplinary. 
But some studies focus on either side, like the viewpoint of audio engineering \cite{aes1}, 
human factors \cite{hf1,hf2}, psychoacoustics \cite{pampas,psysoni}, design \cite{designp} or arts \cite{arts1,arts2,artdesignphilosophy}. Readers of such studies can expect to find the terminology from the respective discipline and results that are relevant to the specific discipline. This can be considered a weak interdisciplinarity from the viewpoint of the specific discipline. The same is partly true for sonification studies that focus on solving a specific problem, like \cite{vocaleeg,cars}. Here, the data and the problem come from one discipline, the \emph{domain science}, while the sonification to solve the problem comes from another. The contributions of each discipline are clearly separated. This makes the collaboration easier. This seems like a good example for the successful determination of identity criteria: the data comes from one discipline, the sonification from another. Nevertheless, it may be difficult for a sonification researcher to read various papers like this, as the terminologies used to describe the sound may vary across studies, i.e., across involved disciplines. This is because sonification research lacks a common terminology, especially concerning sound \cite{shintro,intson1,doom}. Moreover, sonification researchers are often interested in sonification from a multitude of viewpoints and, thus, strong interdisciplinarity. Examples for interdisciplinary discourse on sonification were already provided in the introduction, like \cite{sonihb,davidbuch}.

But even these prominent examples of interdisciplinary sonification research sometimes suffer from the strong interdisciplinarity problem. In a short passage of his book, Worrall mixes terminologies from multiple disciplines, writing that ``Mappings were to pitch, amplitude and frequency modulation (pulsing and detuning), filter coefficients (brightness) and onset time (attack).'' \cite[p. 217]{davidbuch} Here, some terms come from the field of audio engineering (amplitude and pitch modulation, filter coefficients, attack time) while others come from the field of perception (pitch and brightness). As a consequence, it is neither completely clear how the sound is manipulated (audio engineering) nor how it may sound (psychoacoustics). In another part of the book, he writes about ``pitch, loudness, stereo-location'' \cite[p. 218]{davidbuch}. Clearly, pitch and loudness are perceptual terms coming from the field of psychoacoustics. However, the term ``stereo-location'' is a bit ambiguous, probably referring to both the perceived source location, but also to panning technology in a stereo loudspeaker setup (audio engineering). Even though this term appears to be an audio engineering term, it does not reveal how the stereo location was implemented (amplitude based panning, like the sine law, tangent law, or Chowning's panning law, time-based panning, or fading-based panning \cite[Sects. 7.2 and 9.3]{book}). In these passages, the lack of clear identity criteria is evident: Does the description contribute to a) the reproducibility of sonification, or b) the interpretability? Authors need to identify and clearly separate these two identities of a single sonification.



\subsubsection{Communication Discrepancies}
Communication discrepancies easily arise when talking about sonification. As discussed above, sonification needs to be a) replicable and b) interpretable, which is why we need a physical and a perceptual description of each sonification. A common complaint is that authors do not distinguish between physical and perceptual terms \cite[p. 64]{neuortho}\cite[p. 1077]{acta}\cite[p. 6]{mappingreview}\cite[p.20 and chap. 2]{sonic}. Prominent examples can be found when reading different chapters of The Sonification Handbook \cite{sonihb}: For example, \cite{usepsycho} speak of ``(\ldots) acoustic variables, such as pitch or loudness'' \cite[p. 18]{usepsycho} even though these are clearly auditory (from a perceptual domain), and not acoustic (from the physical domain) variables. 
It seems that the authors use these terms colloquially, and not in the strict sense of acoustics and psychoacoustics. Likewise, \cite[Sect. ``Zipper Noise'']{dict} writes about ``signal parameter such as loudness''. David Worrall writes ``[T]hat we (at least in English) so frequently substitute the word ‘note’ for ‘tone’, and ‘music’ for ‘score’'' \cite[p. 18]{davidbuch}. Some studies only provide sonification descriptions from either side: ``(\ldots) in a sonification of real-time financial data Janata and Childs (2004) used rising and falling pitch to represent the change in price of a stock and loudness to indicate when the stock price was approaching a pre-determined target (\ldots)'' \cite[p. 64]{neuortho} Obviously, this description of the sonification helps the reader imagine how it may sound. But it does not enable the reader to reproduce it. Another example can be found in Worrall's book: ``Earcons are made by transforming a tone’s psychophysical parameters--pitch, loudness, duration and timbre--into structured, non-verbal ‘message’ combinations.'' \cite[Sect. 2.2.1.3]{davidbuch} This citation shows how an author successfully managed to describe earcons on a perceptual level with psychoacoustic terminology, avoiding a mixture with musical or physical terms. Other studies aim at illuminating both the reproduction and the interpretation of a sonification:  \cite{icad2019} give the complete formula
\begin{equation}
\begin{split}
a_\text{out} (\Delta x, \Delta y, \Delta z, t) = g(\Delta y, t) \sum_{n=0}^{N}\left[A(\phi_n (\Delta x,t),\Delta y, \Delta z) \right.\\ 
\left. \times \cos \left[\omega_\text{car}(\phi_n(\Delta x,t)) t + \beta (\Delta z) \cos (\omega_\text{mod} t)\right]\right]  
\end{split}
\end{equation}
and then provide equations for all variables, enabling the reader to reproduce the sonification when the three input variables $\Delta x$, $\Delta y$ and $\Delta z$ are given. In the section titled \emph{Psychoacoustics}, they describe how the signal processing affects perception, using the psychoacoustical terms \emph{pitch} \emph{height}, \emph{chroma}, \emph{loudness}, \emph{brightness}, \emph{roughness}, \emph{fullnes}, \emph{subjective duration}, \emph{tonalness}, and \emph{harmonicity}, and a reference to \emph{auditory scene analysis}. This paper stringently uses psychoacoustic terminology to describe the sonification perception, and signal processing formulas to describe the implementation. What is missing, is the explanation of the formula in audio engineering terms.

\subsubsection{Interdisciplinary Semantic Drifts}
``The vast majority of the tools and techniques used for the computer synthesis of sound have been developed by composers and engineers engaged in the task of making new music'' \cite[p. 46]{davidbuch}. This circumstance is one of the reasons of interdisciplinary semantic drifts in sound terminology. A typical case of a semantic drift can be found in the introductory explanation chapters of Chowning's and Bristow's book \emph{FM Theory \& Applications. By Musicians for Musicians} \cite{chown}. As the book title implies, the inventor and implementor of FM synthesis and the Yamaha DX7 explain their audio invention to musicians. This leads to referring to \emph{frequency modulation} as \emph{vibrato}, to \emph{modulation depth} as \emph{vibrato depth}, and to \emph{modulation frequency} as \emph{vibrato rate}. This aim of translating from the audio domain to the musical domain for a better understanding may be one of the origins of semantic drifts. Note that, originally, \emph{vibrato} is a musical term. Like many musical terms, it is related to the production mechanism of this characteristic sound: vibrating your finger on the string that you are playing, or vibrating the telescoping slide of a trombone back and forth. And what this does to the sound is more than a frequency modulation. As the frequency envelope of string instruments is not flat, a vibrato is always a combination of a coherent frequency modulation and incoherent (individual for all partials) amplitude modulation. Generally speaking, the inclusion of electrical engineering and recording studio technology in the creative process of music making initiated reconsideration of traditional musical terms, and introduced signal processing terms to the music domain.

Likewise, the term \emph{pitch} from the music domain underwent a semantic drift in the psychoacoustic domain. In the music domain, pitch refers to the vertical dimension of notes, and to musical intervals. It often, but not always, refers to the fundamental frequency. In psychoacoustics, pitch refers to a multidimensional perception consisting of a height dimension, a chroma dimension and a strength that are affected by periodicity, (fundamental) frequency and frequency distribution, sound pressure level, the place-principle and phase locking of neurons in the auditory nerve \cite{schneiderpitch}.



\section{SOLUTION FOR SONIFICATION RESEARCH}
\label{solution}
The examples above show that many general issues of interdisciplinary research apply to sonification research. Consequently, I aim at transferring the proposed solutions to the sonification domain.

\subsection{Name Concepts, Objectives and Practices}
The suggested solution for conflicting paradigms is to name concepts, objectives and practices. This solution can be applied directly on sonification research. Interdisciplinarity within a single sonification study means that different disciplines may be involved in the three aspects of sonification:
\begin{enumerate}
    \item Sound Design Concept
    \item Objective
    \item Method
\end{enumerate}
The sound design concept typically has one source of inspiration, but also two requirements, which are a) reproducibility and b) interpretability \cite{report,arbitrary,taxonomy}. The source of inspiration can come, e.g., from the field of music \cite{musicinfo}, psychoacoustics \cite{schwarzziemer}, design \cite{uisurvey}, speech \cite{vowel}, Gestalt psychology \cite{asasonifi}, or signal processing \cite{icad2019}. The objective of a sonification study can be, e.g, the conceptualization of a sonification design principle \cite{musicmap}, to set up a framework for sonification design \cite{sonipy} or evaluation \cite{nagelsonex}, to evaluate the perception of a sonification \cite{123channels2}, the performance \cite{acta} or user experience \cite{trustus} of sonification users, or the sonification's impact on the individual \cite{grasping}, the society \cite{aware2}, the environment \cite{reef}, etc. While some of these objectives are irrelevant for some disciplines, sonification researchers should always be interested in all these objects of interest, for the bigger picture of sonification. Depending on the objective, different methods may be appropriate. Methods can come from the field of psychoacoustics \cite{pampas}, human factors \cite{jmui}, UX-Design \cite{buzz}, experimental psychology \cite{123channels2}, Gamification \cite{curat}, Human-Computer Interaction (HCI) \cite{multiexperimentshort} and others \cite{evaluation}.

\textbf{Recommendation}: Authors should always refer to all three aspects of sonification, namely 1) Sound Design Concept, 2) Objective, and 3) Method. Authors should always explain which discipline is involved in which aspect. The three aspects and each involved discipline should be mentioned already in the title and abstract of a manuscript.

\subsection{Identity Criteria}
To solve the problem of strong interdisciplinary, the literature suggests that disciplines organize themselves through identity criteria. In the field of sonification, the above-mentioned solution already partially solves this issue: Authors explicitly associate each involved discipline with either of the three aspects of sonification, namely the sound design concept, the objective and/or the method. This assignment is necessary to reveal, whether the sound design concept is musical 
\cite[p. 79]{neuortho}, -- such as using musical scales, reverse scoring, step-sequencer, voice leading or counterpoint -- or whether the objective is musical, i.e., an enjoyable listening experience \cite{nomusic}. In addition, as discussed above, sonification needs to be a) replicable and b) interpretable, which is why we need a physical and a perceptual description of each sonification. In their parameter mapping sonification review \cite{mappingreview}, Dubus and Bresin already expressed the need for a ``structure for the classification of both physical and auditory dimensions`''. In other words:

\textbf{Recommendation:} Authors should always reflect on their sonification's  a) reproducibility using technical/physical terms and b) interpretability using perceptual terms. These different viewpoints should never be mixed up.

\subsection{Lexical Study}
According to the literature on interdisciplinary research, the problem of communication discrepancies can be solved through lexical study. This is important, because the same term can have different meanings or connotations in different disciplines. Several sonification studies already aimed at finding a consistent taxonomy, e.g., concerning design \cite{vistax,termivison}, listening modes \cite{listmode}, interaction \cite{taxint} and auditory display \cite{taxonomy,usepsycho}. other sonification studies have already expressed the need for a consistent sound taxonomy, too\cite{shintro,designp}.

As discussed above, the sound design concept of a sonification can originate, e.g., in musical, psychoacoustical, or Gestalt-psychological considerations and techniques. Consequently, it seems nonsensical to establish a standardized sonification terminology, as it would, e.g., force authors to use a non-musical language to express their musical sonification sound design concept. The same is true for the objective and the method of the sonification study. However, the situation is different concerning reproducibility and interpretability of sonification. The first clearly requires a technical terminology, e.g., from the field of physics/audio engineering \cite{mappingreview}, while the second requires perceptual terms. Here, psychoacoustic terminology has been suggested \cite{usepsycho,walkerkramerpsy}. In their review paper about parameter mapping sonification \cite{mappingreview}, Dubus and Bresin suggest five perceptual high-level categories \emph{Loudness-related}, \emph{Pitch-related}, \emph{Spatial}, \emph{Temporal} and \emph{Timbral} that can serve to describe sonifications. Following this categorization, I carried out a comprehensive lexical study of sound descriptors in the field of audio engineering and psychoacoustics. These can be found in the appendix.

\textbf{Recommendation:} Authors should describe sonifications using audio engineering terminology for the sake of a) reproducibility and using psychoacoustic terminology for the sake of b) interpretability. A comprehensive list of terms, definitions and further readings can be found in the appendix.





\subsection{Name Disciplines, whose terminology you use}
All the above-mentioned solutions underlined the need to name involved disciplines, ideally already in the title and abstract, and assign them to the sound design concept, the objective, or the methodology. I also identified that the requirement of sonifications to be reproducible and interpretable can be covered by the disciplines audio engineering and psychoacoustics, and their respective terminology. However, these are just the most essential disciplines involved.

Whenever you use the terminology from other disciplines, such as music(ology), computer science, cognitive sciences, or alike, say so, before you use them. And state why you use them. 
In his book \emph{Sonification Design}, David Worrall describes that the inclusion of algorithmic music and data-driven compositions in sonification is unfortunate, ``because it blurs purposeful distinctions'' \cite[Sect. 1.9]{davidbuch}. It is crucial that authors clearly make those distinctions.

\textbf{Recommendation:} Explicitly name disciplines whose terminology you use, and whether you use it to describe the sound design concept, to ensure reproducibility of sonification, to describe interpretation/perception of the sonification, to reveal the objective, or the method that you use.

The proposed solution has been applied in some sonification studies. For example, \cite{neuortho} writes: ``(\ldots) a primary physical characteristic of a tone is its fundamental frequency [usually measured in cycles per second or Hz]. The perceptual dimension that corresponds principally to the physical dimension of frequency is 'pitch', or the apparent 'highness' or 'lowness' of a tone.'' \cite[p. 64]{neuortho}. This statement makes the distinction between two disciplines clear, based on their viewpoints and terminologies.

\section{DISCUSSION}
\label{discussion}
Naturally, the vocabulary of a single discipline is limited. Sometimes, authors need to include the terminology from a second discipline, e.g., to ensure a) reproducibility and b) interpretability of a sonification. Theoretically, a sonification does not require signal processing. A famous example is the sonificaton of a ball's velocity when rolling down a ramp \cite[p. 267]{sonifilab}: When placing gut frets or light bells equidistantly above the ramp, such that the ball strikes them when passing, you can hear the ball speed up, because the frequency of bell strikes rises, i.e., the inter-onset interval reduces. Here, the audio engineering terminology needs to be expanded by physical terms to assure reproducibility of the setup. Similarly, the psychoacoustic terminology is not capable of expressing all aspects of sound perception. Here, terms from Gestalt psychology (auditory scene analysis) or soundscape studies may be borrowed. However, this should always be mentioned explicitly. Note that the two problems of sonification research tackled in this study are only the tip of the iceberg. As discussed above, the interdisciplinarity in sonification research goes far beyond sound terminology. However, especially the division of sonification into the three aspects of sound design concept, objective, and methodology should help authors structure and illuminate their work in such way that readers from various disciplines should be able to understand it.



\section{CONCLUSION}
\label{conclusion}
In this paper, we have identified two problems in sonification research related to the use of sound terminology, namely 1.) the lack of a common sound terminology, and 2.) the blurred separation between disciplines' methods and goals. Through consultation of literature on interdisciplinary research, I could relate these two sonification-specific problems to four general problems of interdisciplinary research. Then, I applied the proposed solutions to sonification research, resulting in recommendations concerning sound terminology and the separation of involved disciplines: Authors should not only name all disciplines involved in their study, but also assign them to the aspect 1.) Sound Design Concept, 2.) Objective, and/or 3.) Method. Regarding the requirement of a sonification to be a) reproducible and b) interpretable, authors should describe both viewpoints of their sonification, ideally using audio engineering and psychoacoustic terminology, respectively. They should consistently stick to the terminology of the respective discipline for the respective aspect. The lexical study provided in the appendix helps sonification researchers understand the sound vocabulary of audio engineering and psychoacoustics, avoid common misconceptions, and find further reading.
\bibliographystyle{unsrt}  
\bibliography{arx.bib}  

\section*{APPENDIX}
What follows is a comprehensive list of sound terms from the audio engineering domain, and the perception/psychoacoustics domain, subdivided into the six high-level categories \emph{Loudness-Related}, \emph{Pitch-Related}, \emph{Spatial}, \emph{Temporal}, \emph{Timbral} and \emph{Other}.
\subsection{Audio Engineering Terms}
Several textbooks and chapters provide a comprehensive overview about acoustics, audio engineering, and signal processing, like \cite{dict,chown,hartmann,roederer,gins1,gins2,asara,fili,current,mores,osc,heller}.
\subsubsection{[Audio Engineering] Loudness-Related}\noindent\textbf{Amplitude} is often referred to as the strength of a signal regardless of its frequency, and is measured in decibels of sound pressure level \cite[sect. 1.1.1]{gins1}\cite[p. 30]{chown}\cite[Sect. “Amplitude”]{dict}, but it can also refer to particle elongation/displacement, velocity or acceleration, and to voltage rather than power \cite[Sect. “Amplitude”]{dict} \cite[sect. 1.1.5.]{brun}. \\\textbf{Amplitude modulation} \cite[Sect. “Amplitude Modulation, abbr. AM”]{dict}. \\\textbf{Attenuation} is the amplitude reduction, often measured in decibels of sound pressure level \cite[Sect. “Attenuation”]{dict}. An \\\textbf{Attenuator} aka. \emph{Pad} or \emph{Loss Pad} is a device to reduce the signal amplitude  \cite[Sects. “Attenuator”, ``Pad'']{dict}\cite[p. 149]{musictech}
. \\\textbf{Beats} are periodic envelope fluctuations resulting from superposition of waves with similar frequency, which can be produced, e.g., through additive synthesis or amplitude modulation\cite[p. 9]{asara}. \\\textbf{Complex amplitude} is a complex number to modify signal amplitude and phase \cite[sect. 2.2]{dey} \cite[sect. 1.1.5.]{brun}. \\\textbf{Damping} is the addition of friction to a resonant system, reducing the magnitude of the resonance. Resistance is the electrical analogue to damping; not to be confused with dampening, which means adding water. \cite[Sects. ``Damp, Damping’’ and ``Dampen’’]{dict}. \\\textbf{Decibel} (dB) is a unitless, logarithmic measure of power, like sound intensity, but it is also used to quantify signal strength in terms of squared sound pressure, voltage, particle elongation, velocity (aka. velocity level), acceleration, or alike \cite[Sect. ``Decibel, or dB’’]{dict}\cite[p. 11]{asara}\cite[Sect. 2.10]{heller}. \\\textbf{Dry} refers to a signal with little reverberation \cite[Sect. ``Dry’’]{dict}. When the signal also contains little ambient noise, it is called \emph{Dead} \cite[Sect. ``Dead’’]{dict}\cite[p. 48]{musictech}. \textbf[Displacement] or \emph{Deflection/Elongation} over time (not amplitude over time) is plotted in audio files in time domain and is proportional to voltage \cite{seng}. \\\textbf{Dynamic range} can refer to the ratio of the loudest to the softest part, measured in dB, but also to the signal-to-noise ratio \cite[Sect. `` Dynamic Range’’]{dict}. \\\textbf{Fade} is the gradual change of volume, like fading in or our \cite[Sects. ``Fade’’, `` Fade In’’ and ``Fade Out’’]{dict}\cite[p. 71]{musictech}. \\\textbf{Gain} is the amount of amplification and can be considered a multiplication of sound pressure by a real value \cite[Sects. ``Amplifier'', ``Gain’’, ``Voltage Gain'']{dict}. \\\textbf{Inverse Distance Law} describes that the sound pressure of a point source decays with a factor $p \propto 1/r$, where $r$ is the distance from the source. The \emph{Inverse Square Law} describes the same circumstance, but refers to sound intensity $I \propto 1/r^2$ \cite[p. 576]{schneiderfunda}\cite[Sect. ``Inverse Square Law’’]{dict}\cite[Sect. 2.2]{heller}\cite[p. 99]{musictech}. \\\textbf{Master} is a gain control for all output channels at once \cite[Sect. ``Master’’]{dict}. \\\textbf{Mute} means to silence a track or a channel \cite[Sect. ``Mute’’]{dict}. \\\textbf{Particle Displacement} is proportional to voltage in audio signals\cite[sect. 1.1.5.]{brun}. \\\textbf{Normalization} means increasing the elongation/deflection/voltage/sound pressure such that the peak equals the largest possible value \cite[Sect. ``Normalize’’]{dict}\cite[p. 137]{musictech}. \\\textbf{Particle Velocity} is the time derivative of particle displacement \cite[sect. 1.1.5.]{brun}. \\\textbf{Peak} is the largest absolute value of an audio signal over a considered time interval \cite[Sect. ``Peak’’]{dict}\cite[p. 154]{musictech}. \\\textbf{Resonances} are eigenfrequencies of a physical system in which it will vibrate stronger and longer, when excited with a signal containing these frequencies \cite[Sect. ``Resonance’’]{dict}. \\\textbf{Root Mean Square} of sound pressure or voltage quantifies the average volume \cite[Sects. ``RMS’’, "Volt, Voltage'']{dict}. \\\textbf{Sound energy} aka. acoustic energy in W is the amount of kinetic and potential energy \cite[p. 3640]{gins2}, i.e., ``variation of energy produced by the acoustic perturbation'' \cite[sect. 1.2.5.]{fili}. \\\textbf{Power}, is the rate of doing work, measured in Watts [W] and equals voltage times current \cite[Sects. ``Power'', ``Watt'']{dict}. \\\textbf{Sound intensity} $I$ in $\mathrm{W}/\mathrm{m}^2$ \cite[Sect. ``Intensity’’]{dict} is the power per area carried by the wave. It is the product of sound pressure and particle velocity \cite[pp. 572f]{gins2} and proportional to the squared sound pressure \cite[sect. 1.1.5.]{brun}\cite[sect. 1.2.5.]{fili}\cite[p. 568]{schneiderfunda}\cite[p. 12]{asara}. 
 \\\textbf{Sound Power} in Watts cannot be measured directly and tends to have a complicated relationship to sound pressure \cite[Sect. ``Sound Power'']{dict}. \\\textbf{Sound pressure} $p$ is the ``magnitude of the pressure variation'' \cite[p. 18]{brun}. \\\textbf{Sound pressure level}  \cite[sect. 1.3.1]{gins1}. Weighted sound pressure levels in $\text{dB}_\text{A to C}$ attenuate certain frequencies before calculating the sound pressure level \cite[sect. 1.3.2]{gins1} \cite[Sects. ``dBA’’ and “A-Weighting’’]{dict}.

\subsubsection{[Audio Engineering] Pitch-Related}
\noindent\textbf{Angular Frequency} $\omega=2\pi f$ is the circumference of the unit cycle times the frequency. A \\\textbf{Cent} is the $\sqrt[1200]{2}$ and divides the octave into $1200$ equal frequency ratios \cite[Sect. “Cents”]{dict}\cite[p. 31]{musictech}. \\\textbf{Complex Tone} is a sound that contains discrete (not continuous)  spectral peaks\cite{schneid} \cite[p. 24]{asara}.\\\textbf{Decade} refers to a frequency ratio of $10:1$ and is sometimes used to describe the rolloff of a filter, like $20$ dB per decade \cite[Sect. ``Decade’’]{dict}. \\\textbf{Frequency} in Hertz (Hz) which replaced the earlier unit of cycles per second (cps) is the number of exact repetitions per second \cite[Sects. ``Cycles per Second, cps’’, ``Frequency’’ and `` Hertz, Hz’’]{dict}\cite[pp. 78, 87]{musictech}. \\\textbf{Frequency Modulation} means processing frequencies so that they vary continuously over time \cite[p. 10]{asara}\cite[p. 78]{musictech}. \\\textbf{Fundamental Frequency} is the lowest frequency from a complex tone \cite[p. 43]{gins1} \cite[Sect. `` Fundamental’’]{dict}\cite[p. 8]{asara}\cite[p. 80]{musictech}. \\\textbf{Period} is the reciprocal of frequency and can refer to the duration until a function repeats itself\cite[sect. 1.1.1]{gins1}\cite[p. 7]{asara}. \\\textbf{Periodicity} is the frequency of exact signal repetition, irrespective of the fundamental frequency \cite[p. 43]{gins1}. \\\textbf{Pitch    Bend} is a controller and/or MIDI message to alter the speeed of a sound playback, simulating the musical playing technique of pitch bending, known., e.g., from guitars, violins and trombones \cite[p. 158]{musictech}. \\\textbf{Pure tone} aka. \emph{Simple Tone} has a sinusoidal waveform and contains only one frequency \cite[Sect. `Pure Tone'']{dict}\cite[p. 24]{asara}. \\\textbf{Sweep} aka. \emph{Chirp} is a frequency that continuously rises (or falls) as a mostly exponential function of time \cite[p. 382]{dict}. \\\textbf{Wow} is a slow (below $5$ Hz) frequency modulation caused by speed variation in tape recorders \cite[Sect. ``Flutter’’]{dict} 

\subsubsection{[Audio Engineering] Spatial}
\noindent\textbf{Amplitude-Based panning} means gain weighting between several channels according, e.g., to Chowning's/sine/tangent panning law, Vector Base Amplitude Panning (VBAP) or Multiple Direction Amplitude Panning (MDAP), or through microphone techniques like Y-X, ORTF, Blumlein aka. intensity stereo \cite[Sects. `` Intensity Stereo’’, ``X-Y Stereo’’]{dict}\cite{current}). \\\textbf{Depth} refers to aims at manipulating the apparent distance of a sound source in stereo, and thus, indirectly, its size \cite[Sect. ``Depth’’]{dict}. \\\textbf{Dichotic}, \\\textbf{Diotic} and \emph{Monotic} presentation refers to headphone signal being individual for each ear, equal for each ear, and presented to one ear only \cite[Sects. ``Dichotic’’, ``Diotic’’ and ``Monotic’’]{dict}. \\\textbf{Sonic Environment} refers to sounds at a certain place and time \cite{soundecol}. \\\textbf{Phantom Source} are desired location and extent of an auditory event in stereo setups \cite[p. 155]{musictech}.

\subsubsection{[Audio Engineering] Temporal}
\noindent\textbf{Burst} is a test signal that typically lasts some milliseconds, like a tone burst and a noise burst \cite[Sect. ``Burst'']{dict}. \\\textbf{Cycle} includes all elements that repeat periodically \cite[p. 7]{asara}. \\\textbf{Echo} is a distinct, damped repetition of a signal with a delay of $50$ ms or more \cite[Sect. ``Echo’’]{dict}\cite[p. 10]{asara}. \\\textbf{Impulse} is a short, broadband spike \cite[p. 96]{musictech}.\\\textbf{Initial Phase} ($\phi_0$) is the starting point of an oscillation function $p(t)=\hat{A} \sin(\omega t + \phi_0)$ \cite{acta}. \\\textbf{Inter-Onset-Interval} (IOI) is the duration between two successive note-, noise-, impulse-, or sound-onsets \cite[p. 4]{london}. \\\textbf{Period} is the duration of a full cycle of a waveform \cite[p. 154]{musictech}. \\\textbf{Phase} can describe the argument of an oscillation function or the angle of a single frequency's complex amplitude \cite[sect. 2.2]{dey}\cite[Sect. 5.1.4]{book}. \\\textbf{Transient} is a nonrepeating waveform, including note onsets, offsets, decays and modulation \cite[Sect. ``Transient’’]{dict}.

\subsubsection{[Audio Engineering] Timbral}
\noindent\textbf{Bandwidth} is the frequency region in which a spectrum contains significant energy \cite[sect. 1.5.1]{gins1}\cite[Sect. ``Bandwidth'']{dict}. \\\textbf{Bass} is often considered the low-frequency portion of the audible frequency range up to about $200$ Hz\cite[Sect. ``Bass'']{dict}. \\\textbf{Broadband} or \emph{Wideband} refers to a wide distribution of spectral energy \cite[Sects. ``Broadband’’, ``Wideband’’]{dict}. \\\textbf{Clipping} means that the waveform looks like clipped by a pair of scissors, producing harmonic distortion and adding to the perception of roughness, but may also make speech easier to understand in noisy environments \cite[Sect. ``Clipping’’]{dict}. \\\textbf{Corner Frequency} is the frequency of a filter where the amplitude is reduced by $3$ dB, like the \emph{Cutoff-Frequency} of a high-pass or low-pass filter \cite[Sects. ``Corner Frequency’’ and ``Cutoff Frequency’’]{dict}\cite[pp. 41, 44]{musictech}. \\\textbf{Duty Cycle} is the pulse width per period in pulse width modulation synthesis, affecting the weighting of harmonics \cite[p. 16]{osc}\cite[p. 59]{musictech}. \\\textbf{Extreme highs} are frequencies above $10$ kHz, while \\\textbf{Extreme Lows} are frequencies below $40$ Hz \cite[Sect. ``Distortion’’]{dict}. \\\textbf{Filters} include \emph{High}, \emph{Low}, \emph{Stopband/Band Reject/Band Elimination} and \emph{Band Pass Filters} that let frequencies above, below, outside and, respectively, between the corner frequencies pass while attenuating the others. \emph{Notch Filters} attenuate, \emph{Resonance/Peak Filters}  amplify mostly a narrow frequency region. \\\textbf{Comb Filters} exhibit a series of deep notches. \emph{Shelf Filters} have a flat response over one or two large frequency ranges and a gradual slope within the transition range.\cite[Sects. ``Bandpass Filter’’, “Band Reject Filter'', ``Brickwall Filter and Low-Pass Filter'', ``Comb Filter’’, ``High-Pass Filter’’, ``Notch Filter'', ``Shelving’’]{dict}\cite[p. 18]{asara}\cite{mores}\cite[pp. 16, 35, 74, 88, 113f, 138f, 154]{musictech}. \\\textbf{Frequency bands}, like octave bands and 1/3 octave bands group the frequency spectrum into smaller portions \cite[sect. 1.3.3.]{gins1} \cite[Sect. ``One-third Octave Filter’’]{dict}. \\\textbf{Harmonic Series} are frequencies that are integer multiples of a fundamental frequency\cite[p. 86]{musictech}.\\\textbf{Highs} are frequencies between $4,000$ and $10,000$ Hz \cite[Sect. ``Distortion’’]{dict}\cite[p. 88]{musictech}. \\\textbf{Lows} are frequencies roughly between $40$ and $300$ Hz \cite[Sect. ``Distortion’’]{dict}. \\\textbf{Mid-Bass} means frequencies around $200$ to $300$ Hz \cite[Sect. ``Bass'']{dict}. \\\textbf{Mid-range} are frequencies between $300$ and $4,000$ Hz \cite[Sect. ``Midrange’’]{dict}. \\\textbf{Modulation Index} is a ratio that expresses the amplitude ratio of carrier and modulation signals in AM synthesis and the frequency deviation between carrier and the highest or lowest sideband in FM synthesis \cite[Sect. ``Modulation Index’’]{dict}. \\\textbf{Narrowband} means that most spectral energy is concentrated in a narrow frequency band, e.g. $400$ cent \cite[Sect. ``Narrowband'']{dict}. \\\textbf{Noise} can mean an unwanted sound not related to the desired signal, or a stochastic, aperiodic process, like \emph{Gray, White, Brown/Red, Pink, Blue, Purple Noise} \cite[Sects. ``Noise’’, ``One-over-f (1⁄f ) Noise’’, ``White Noise'']{dict}\cite[sects. 1.5.1f]{gins1}. \\\textbf{Overdrive} means that the gain drives the signal above the linear operating level of equipment, causing overload distortion \cite[Sects. ``Operating level’’, ``Overdrive’’ and ``Overload’’]{dict}. \\\textbf{Passband} is the frequency region not rejected by a filter \cite[Sect. ``Passband’’]{dict}\cite[pp. 151f]{musictech}. \\\textbf{Pulse Width Modulation} (PWM) is a technique to produce complex tones \cite[Chap. 2]{osc}.  \\\textbf{Q}, \emph{Q-} or \emph{Quality-Factor} is the steepness of a filter in frequency domain \cite[Sect. ``Q’’]{dict}\cite{mores}. \\\textbf{Slope} (aka. skirt) is the steepness of a filter in frequency domain, expressed in dB per decade or octave \cite[Sect. ``Slope’’]{dict}. \\\textbf{Sound Synthesis} includes techniques such as \emph{Additive/Fourier Synthesis}, \emph{Frequency Modulation (FM) Synthesis}, \emph{Granular Synthesis}, \emph{Ring Modulation}/\emph{Amplitude Modulation Synthesis}, \emph{Physical Modeling}, \emph{Subtractive Synthesis} and combinations \cite[Sects. ``Additive Synthesis'', ``FM Synthesis’’, ``Ring Modulator'', ``Subtractive Synthesis’’]{dict}\cite[Sect. 6.1]{guil}\cite{chown}\cite[pp. 4, 7, 76f, 82, 156f]{musictech}\cite[chap. 9]{hartmann}. \\\textbf{Spectral Centroid} is the center of gravity of a magnitude spectrum \cite[Sect. 4.3.1]{current}. \\\textbf{Spectrum} or \emph{Frequency Spectrum} is the frequency representation of a time signal, indicating amplitude and phase per frequency over a time frame \cite[Sect. ``Spectrum’’]{dict}\cite[p. 8]{asara}. \\\textbf{Synthesizer} is a hard- and/or software that electronically produces sound \cite[Sect. ``Synthesizer’’]{dict}. \\\textbf{Treble} are frequencies above about 2 kHz \cite[Sect. ``Treble’’]{dict}. \\\textbf{Waveform} is the shape of an oscillation over time, like \emph{Sine}, \emph{Triangle}, \emph{Sawtooth}, \emph{Square}, \emph{Pulse rain}, \emph{Complex Waves} and \emph{White} (random) \emph{Noise} \cite[Sects. ``Complex Wave’’, ``Sine Wave’’, ``Square Wave’’, ``Waveform’’]{dict}. \\\textbf{(Temporal) Envelope} is the amplitude (not displacement/deflection/elongation/voltage) over time. It can also refer to an amplitude or gain function $A(t)$, like \emph{Attack Decay Sustain Release (ADSR)} curves of synthesizers 
\cite{schneid} \cite[Sect. “Additive Synthesis”]{dict}\cite[pp. 4, 66f]{musictech}\cite{soundecol}\cite[p. 595]{schneiderfunda}.

\subsubsection{[Audio Engineering] Other}
\noindent\textbf{Audio Effects} is signal processing to modify an audio signal, like \emph{Chorus}, \emph{Compressor}, \emph{Delay}, \emph{Distortion}, \emph{Equalizer}, \emph{Filter}, \emph{Flanger}, \emph{Phaser}, \emph{Reverb} \emph{Wah Wah} \cite[Sects. ``Chorus’’, ``Effects’’, ``Equalizer'', ``Flanging’’, `` FX’’]{dict}\cite[sect. 6.3.1. and p. 169]{guil}\cite[pp.  32, 37, 41, 50, 67, 156]{musictech}. \\\textbf{Carrier} is a signal about to be modulated by another signal, the \emph{modulator}, e.g., in AM and FM synthesis \cite[p. 29]{musictech}. \\\textbf{Low Frequency Oscillator} (LFO) has a (fundamental) frequency way below $20$ Hz and is used to modulate audio material \cite[p. 110]{musictech}. \\\textbf{Musical Instrument Digital Interface} (MIDI) is a protocol to transmit control data between audio (and lighting) equipment \cite[pp. 122ff]{musictech}. \\\textbf{Open Sound Control} (OSC) is a network protocol for communication between multimedia devices \cite[p. 145]{musictech}\cite[Chap. 8]{osc}.

\subsection{Psychoacoustic Terms}
Several textbooks and chapters provide a comprehensive insight into psychoaocustic terminology, methods and paradigms, like \cite{book,schneiderpitch,roederer,current,heller,schneiderfunda,schneid,fastl,hearing,blauerte}.
\subsubsection{[Psychoacoustics] Loudness-Related} 
\noindent\textbf{Fluctuation Strength} refers to the intensity of loudness fluctuation caused, e.g., by amplitude or frequency modulation or beats. The unit of fluctuation strength is vacil, but it is also often given in percent\cite[chap. 10]{fastl}\cite[Sect. ``Beats'']{dict}. \\\textbf{Loudness} is the perception that lets listeners order sounds from soft to loud, i.e., the subjective strength of an audio signal that depends on amplitude and frequency but even on spatial and temporal attributes.\cite[p. 22]{asara}\cite[sect. 1.3.2]{gins1}\cite[Sects. ``Loudness’’ and ``Sone’’]{dict}. \\\textbf{Masking} means that the presence of one sound inhibits our ability to hear another simultaneous (\emph{Simultaneous Masking}), or shortly preceding (\emph{Pre-Masking}) or subsequent (\emph{Post-Masking}) sound. Typically, loud sounds mask soft sounds, low frequencies mask higher frequencies \cite[Sect. ``Masking’’]{dict}\cite[Sect. 4.3]{book}. \\\textbf{Phon} is the unit of loudness level and is a frequency-compensated decibel scale (aka. \emph{Fletcher-Munson-Curve} or \emph{Equal-Ludness-Contour}) but not a unit of loudness \cite[p. 127]{hartmann}\cite[p. 23]{asara}\cite[pp. 76, 156]{musictech}. \\\textbf{Sone} is a measure of loudness, where a $1$ kHz pure tone at $40$ dB equals $1$ Sone, while a sound twice as loud equals $2$ Sone, etc. \cite[p. 22]{asara}. \\\textbf{Specific Loudness} is the loudness per critical frequency band. Their integral is the overall loudness \cite[Chap. 8]{fastl}. \\\textbf{Threshold of Hearing} aka. \emph{Absolute Threshold} or \emph{Minimum Audible Field} (MAF) is the sound pressure level at the eardrum required to be just audible, which is highly frequency dependent \cite[Sects. ``Minimum Audible Field, MAF’’, ``Threshold of Hearing’’]{dict}.

\subsubsection{[Psychoacoustics] Pitch-Related}
\noindent\textbf{Absolute Pitch} aka. \emph{Perfect Pitch} refers to a listeners’ ability to assign sounds their corresponding position or note on a musical scale \cite[Sect. ``Absolute Pitch’’]{dict}. \\\textbf{Combination Tones} aka. \emph{Phantom Tones}/\emph{Tartini Tones} are audible frequencies produced through nonlinearities in our hearing system that are not physically present in the sound, like \emph{Combination Tones} $f_1+f_2$, \emph{Difference Tones} $f_2-f_1$ and \emph{Cubic Difference Tones} $2f_1-f_2$ \cite[Chap. 12]{hearing}\cite[pp. 52, 206f]{musictech}. \\\textbf{(In)Harmonicity} refers to whether simultaneous frequencies seem to exhibit a harmonic series or not, which can affect simultaneous stream segregation and, therefore, pitch, consonance, timbre and tonal fusion, \cite{schneiderpitch}\cite[Sect. 4.5.2]{book}. \\\textbf{Mel} is the unit of pitch, where $1000$ Mel = pitch evoked by a $1$ kHz pure tone at $40$ phon (or dB) \cite[Sect. ``Mel’’]{dict}\cite[p. 22]{asara}. \\\textbf{Missing Fundamental} aka. \emph{Residual Pitch} or \emph{Virtual Pitch} means that the perceived pitch of a sound equals the pitch of a pure tone whose frequency is not present in the sound \cite[Chap. 5]{fastl}\cite{schneiderpitch}\cite[Sect. ``Residue’’]{dict}. \\\textbf{Pitch} is the perception that lets listeners order sounds from low to high (height) and that includes octave identity/equivalence (chroma) \cite{schneiderpitch}\cite[Chap. 5]{fastl}\cite[p. 22]{asara}\cite[Chap. 12]{hearing}. \\\textbf{Pitch Strength} means how distinct a sound’s pitch is. It is, e.g., low for clicks, higher for inharmonic series, and even higher for pure tones in the with midrange frequency and sufficient loudness \cite[Chap. 5]{fastl}\cite{schneiderpitch}. \\\textbf{Shepard Illusion} is pitch perception with clear chroma but ambiguous height and can be created through Shepard tones \cite{schneiderfunda,schneiderpitch}\cite[Sect. 4.5.4.1]{book}.

\subsubsection{[Psychoacoustics] Spatial}
\noindent\textbf{Ambience} refers to acoustic qualities of a listening space, including echoes, reverberation and background noise and may provide a certain auditory ‘atmosphere’ \cite[Sect. ``Ambience'']{dict}. \\\textbf{Apparent Source Width} aka. \emph{Perceived Source Extent} refers to the perceived size of a sound source, sometimes categorical, sometimes indicated in degrees \cite{current}\cite[Chap. 6.2.2]{book}. \\\textbf{Auditory Event Angle/Distance/Location} refers to the perceived source angle/distance/location irrespective of the physical position \cite{current} \cite[Sect. 4.4]{book}\cite{blauerte}.
\\\textbf{Precedence Effect} aka. \emph{Haas Effect} or \emph{Law of the First Wavefront} means that the first wavefront arriving at the ears determines the perceived source location, even in the presence of louder reflections and reverberation \cite[Sect. ``Haas effect’’]{dict}\cite[Sect. 4.4.5]{book}\cite[p. 85]{musictech}. \\\textbf{Soundscape} refers to how the sonic environment is perceived and understood \cite{soundecol}.

\subsubsection{[Psychoacoustics] Temporal}
\noindent\textbf{Rhythm} is a temporal pattern of one auditory stream and somewhat related to fluctuation strength and subjective duration \cite[Chap. 13]{fastl}. \\\textbf{Subjective Duration} seems quite proportional to physical duration above $100$ ms. But temporal masking may affect subjective duration and, e.g., $50$ ms sound longer than one half of $100$ ms\cite[Chap. 12]{fastl}. \\\textbf{Temporal Masking} contains pre-masking of up to $30$ ms, the \emph{Overshoot} phenomenon at masker onsets, simultaneous masking, a short sustain of $5$ ms and post-masking up to $100$ ms \cite[Sect. 4.3.1.5]{fastl}.

\subsubsection{[Psychoacoustics] Timbral}
\noindent\textbf{Brightness} (from warm/dull to bright/shrill) is considered the most prominent feature in timbre evaluation and seems closely related to the spectral centroid of a signal\cite[p. 15]{book}\cite[p. 595]{schneiderfunda}. \\\textbf{Clangorous} is a continuous loud noise of something being hit or rung, often metallic \cite[p. 704]{schneiderpitch}. \\\textbf{Fullness} aka. \emph{Richness}, sometimes \emph{Volume} is mostly affected by the spectral distribution and bandwidth, and can be increased, e.g., by adding or modulating frequencies \cite{schneid,icad2019}. \\\textbf{Noisiness} indicated in Noy refers to perceived noise loudness, not to the degree of tonalness \cite[Sect. ``Noy’’]{dict}. \\\textbf{Percussive} to \emph{Mellow} is the degree of how impulsive the excitation of a source sounds, where hammered/struck often sound more percussive than plucked, and even more than blown/stringed\cite[p. 701]{schneid}\cite[p. 595]{schneiderfunda}. \\\textbf{Roughness} in asper is perceived when two or more than $16$ but less than one critical bandwidth apart \cite[Sect. ``Beats'']{dict}\cite{aureswohl}\cite{schneid}. \\\textbf{Sharpness} is equal or related to \emph{Brightness}, but has the measure acum that can be approximated by psychoacoustic models \cite{aureswohl}\cite[Chap. 9]{fastl}. \\\textbf{Sound Color} aka. \emph{Tone color} refers to the timbre of quasi-stationary sounds, excluding attack transients and decay \cite{schneid}\cite[p. 393]{dict}. \\\textbf{Timbre} includes spectral and short-term temporal aspects of sound other than pitch and loudness \cite{schneid}. \\\textbf{Tonality} aka. \emph{Tonalness} ranges from tonal to noisy and is affected by frequency, harmonicity, periodicity and bandwidth, and considered a parameter of sensory euphony \cite[p. 41]{book}\cite[p. 700 and Sect. 32.2.3]{schneid}\cite{aureswohl}.

\subsubsection{[Psychoacoustics] Other}
\noindent\textbf{Auditory} refers to the psychological representation of sound \cite[Chap. 4]{book}\cite[p. 586]{schneiderfunda}. \\\textbf{Auditory Stream} is the auditory counterpart of a visual object, i.e., aspects of sound that seem to belong together  \cite[Sect. 4.5]{book}\cite[Sect. 32.2.5]{schneid}. \\\textbf{Critical Bandwidth} is a region of $1.3$ mm on the basilar membrane in which neurons collectively process sound.  \cite[Sect. ``Critical Band’’]{dict}\cite[p. 584]{schneiderfunda}. \\\textbf{Just Noticeable Difference} aka. \emph{Difference Limen} is the increment in a stimulus that is just noticed  in a specified fraction of trials \cite[Sect. ``Just Noticeable Difference, JND’’]{dict}\cite[p. 23]{asara}. \\\textbf{Perception} is a psychological impression often resulting from sensation plus cognitive processing in the brain \cite[Sect. 30.1.2]{schneiderfunda}
\\\textbf{Sensation} can be considered an inter-subjective ur-perception before the subjective interference through context, experience, preference, etc. Psychoacoustic models predict sensation and not perception \cite[Chap. 4]{book}\cite[Sect. 1.4]{fastl}. \\\textbf{Sensory Euphony} aka. \emph{Sensory Pleasantness} or \emph{Psychoacoustical Annoyance} tends to be considered as a nonlinear combination of \emph{Loudness}, \emph{Roughness}, \emph{Sharpness} and \emph{Tonality} \cite[Chap. 9]{fastl}. \\\textbf{Sound Object}  is the smallest self-contained element of a soundscape \cite{soundecol}.

\subsection{Danger of Confusion}
I identified terms with different meanings in different disciplines. The following list shall help researchers avoid mistakable expressions:
\\\textbf{A-B Stereo $\neq$ A-B testing $\neq$ ABX Test}, the first refers to a spaced microphone recording technique, the second to the direct comparison of two versions of the same signal and the latter is an experiment in which participants guess whether a test stimulus X was equal to reference stimulus A or B\cite{current}\cite[Sects. “A-B Stereo”, “A-B Testing” and “ABX Test”]{dict}. \\\textbf{
Acoustic} $\neq$ \textbf{Auditory}, the first refers to physics, the second to perception of sound \cite{blauerte}. \\\textbf{Acoustic Recording} $\neq$ \textbf{Sound Recording}, the first refers to recording without the use of electricity, the second to all sound recording \cite[Sect. “Acoustic Recording”]{dict}. \\\textbf{Amplitude Modulation} $\neq$ \textbf{Tremolo}, the first refers to analogue or digital signal processing, the latter to an articulation technique, e.g., of string instruments \cite[pp. 7, 8]{musictech}. \\\textbf{Audio Track} $\neq$ \emph{Audio Channel}. Each audio channel can contain a mixture of signals from one or more audio tracks, which can be recorded and manipulated separately, before routing them to discrete output channels. \cite[Sects. ``Multitrack’’, ``Tracking'']{dict}\cite[pp. 31, 132]{musictech}. \\\textbf{Auditory Stream} $\neq$ \textbf{Audio Stream}. An auditory stream is comparable to a visual object, i.e., it includes sound aspects that are perceived as being part of the same. An audio stream is the broadcast of one or multiple audio channels. \cite[Sect. ``Streaming'']{dict}. \\\textbf{Bandwidth} in terms of the audio signal $\neq$ \textbf{Bandwidth/Throughput/Data Rate} of a network connection or interface \cite[p. 16]{musictech}. \\\textbf{Bar} in physics $\neq$ \textbf{Bar} in music. In physics, Bar is a unit of pressure, and a bar is a uniform elastic rod or beam with irregular eigenfrequencies. The latter is a segment of music often having a uniform duration,  number of beats and structure of accentuation. \\\textbf{Beats} in physics $\neq$ \textbf{Beats} in music. The first refer to envelope fluctuation caused by superposition of similar frequencies, the latter to the pulse, to the rhythm section of a piece of music, or to the instrumental, e.g., of hip-hop music \cite[p. 19]{musictech}. \\\textbf{Current} $\neq$ \textbf{Charge} \cite[Sects. “Ampere, abbr. Amp or A” and Charge]{dict}. While the current is a property of a circuit, the charge flows. There is no such thing as a current flow.
\\\textbf{dB} $\neq $ \textbf{dB}$_\text{SPL}$$ \neq$ \textbf{dBA}. The first is a power level with arbitrary reference, the second uses a reference pressure of $20$ micropascals, just like the third one, which also adds a frequency-weighting \cite[Sects. “A-Weighting”]{dict}\cite[pp. 46, 49]{musictech}. \\\textbf{Depth} $\neq$ \textbf{Auditory Distance} $\neq$ \textbf{Source Distance}, the first refers to stereo methods to affect auditory distance, which is the perceived source distance, while source distance refers to the physical distance of a source \cite[Sect. ``Depth’’]{dict}\cite{blauerte}. \\\textbf{Delay} as an audio effect $\neq$ \textbf{Delay} of a network connection or interface. The first is a single or a number of damped repetitions, the latter is latency of signal transfer \cite[pp. 50, 108]{musictech}. \\\textbf{Elongation/Deflection/Voltage} $\neq$ \textbf{Amplitude}, the first refer to the course of the time signal, the second to the peak of the absolute time signal \cite{seng}. \\\textbf{Frequency/Fundamental Frequency} $\neq$ \textbf{Pitch}, the first is physical, the second is a perceptual quality affected strongly by the first, but also by periodicity and sound pressure level \cite{schneid}. \\\textbf{Frequency Modulation} $\neq$ \textbf{Wow} $\neq$ \textbf{Vibrato}. Frequency modulation describes periodic, continuous frequency variation through signal processing, wow is a frequency modulation resulting from tape speed alterations, and vibrato is a musical instrument playing technique where frequencies and amplitudes of complex tones vary continuously and regularly \cite[Sect. ``Wow'']{dict}\cite[pp. 10, 25]{asara}. \\\textbf{Jitter} in speech $\neq$ \textbf{Jitter} in an audio network connection or interface \cite[Sect. ``Jitter’’]{dict}. \\\textbf{Loudness} $\neq$ \textbf{Volume} $\neq$ \textbf{Sound Pressure Level/Amplitude/Gain}. The first is a perceptual quality, the second is a colloquial term loosely related to the others, the latter refer to physical quantities \cite{schneid}\cite[Sects. ``Amplitude'', ``Level'', ``Voltage Gain'']{dict}\cite[sect. 1.1.1]{gins1}\cite[p. 12]{asara}\cite[pp. 81, 110]{musictech}. \\\textbf{Meter} in music $\neq$ \textbf{Meter} as a measuring device $\neq$ meter as the SI unit of length \cite[Sect. ``Meter’’]{dict}. \\\textbf{Monitor} in audio (aka \emph{Studio Monitor} or \emph{Stage Monitor} $\neq$ \textbf{Monitor} in computer science (visual display) \cite[Sects. ``Monitor’’ and ``Near-Field Monitor’’]{dict}\cite[pp. 127, 133]{musictech}. \\\textbf{Note} $\neq$ \textbf{Tone}. The first is a sign indicating pitch and/or duration of a tone or sound. The second is a sound having a pitch. \cite[p. 25]{asara}. \\\textbf{Overshoot} in audio $\neq$ \textbf{Overshoot} in perception. The first means that equipment like compressors or limiters responds imperfectly to short transients, while the latter means a brief increase of the masking threshold at sound onsets \cite[Sect. ``Overshoot’’]{dict}\cite[Sect. 4.3.1.]{book}. \\\textbf{Polyphonic} in audio engineering $\neq$ \textbf{Polyphonic} in music. The first means the overlap of multiple voice, e.g., in synthesizers and music mixes. The second is a horizontal thinking (like counterpoint), i.e., multiple voices playing melodies that result in consonant intervals on pulse, except when the intention is to create tension \cite[Chap. 5]{roederer}\cite[pp. 711f]{schneid}. \\\textbf{Overtone} $\neq$ \textbf{Harmonic} $\neq$ \textbf{Partial}. The fundamental frequency is usually the lowest frequency and the first partial, and it may be the first harmonic of a spectrum. The first overtone is the second partial. It is also the second harmonic, if the signal is a complex tone, i.e., if all partials are part of a harmonic series. \cite[Sects. ``Overtones’’ and ``Partials’’]{dict}\cite[p. 25]{asara}\cite[pp. 86, 147]{musictech}. \\\textbf{Sharp} in music $\neq$ \textbf{Sharp} in psychoacoustics $\neq$ \emph{Sharp} in audio engineering. In music, sharp is a rise in pitch, e.g. by one half-step. In psychoacoustics, it is a timbral quality, related to brightness. In audio engineering, filters' \emph{Slope}/\emph{Rolloff}/\emph{Attenuation}/\emph{Rejection Rate}/\emph{Q-Factor} are also called sharpness \cite[Sect. ``Sharp’’]{dict}\cite{aureswohl}\cite[p. 15]{hearing}.\cite{schneid}. \\\textbf{Sound Velocity} $\neq$ \textbf{Particle Velocity}, the first is a particle's velocity when moving around its equilibrium position, the second is the propagation speed of a sound wave (perturbation). \\\textbf{Spectral Centroid} $\neq$ \textbf{Brightness}, the first is the center of gravity of a magnitude spectrum, the latter an aspect of timbre perception \cite[Sect. 4.3.1]{current}. \\\textbf{Stereo} sometimes $\neq$ \textbf{Stereo}. While stereo tracks in digital audio workstations often refer to two channels, one for each loudspeaker in a stereo triangle, stereo can also refer to loudspeaker systems using more than two loudspeakers. \cite[Sect. ``Stereophonic'']{dict}\cite[Chap. 7]{current}. \\\textbf{Tone} in physics $\neq$ \textbf{Tone} in music. In physics, tone refers to a pure or complex tone, while in musical terms it can also refer to tone color or even timbre \cite[Sect. ``Tone'']{dict}. \\\textbf{Voice} in speech $\neq$ \textbf{Voice} in music. The first always refert to the human voice, the second can refer to different instruments or registers \cite[pp. 711f]{schneid}. \\\textbf{Timbre} $\neq$ \textbf{Sound Color}/\textbf{Tone Color}. Timbre includes short-term temporal aspects, like initial transients and decays, while tone color/sound color mostly refers to the spectrum of the quasi-stationary part of sounds \cite{schneid}.\\

\end{document}